\documentstyle[12pt]{article}
\catcode`\@=11
\long\def\@makefntext#1{
\protect\noindent \hbox to 3.2pt {\hskip-.9pt
$^{{\ninerm\@thefnmark}}$\hfil}#1\hfill}                
\def\@makefnmark{\hbox to 0pt{$^{\@thefnmark}$\hss}}  
\def\ps@myheadings{\let\@mkboth\@gobbletwo
\def\@oddhead{\hbox{}
\rightmark\hfil\ninerm\thepage}
\def\@oddfoot{}\def\@evenhead{\ninerm\thepage\hfil
\leftmark\hbox{}}\def\@evenfoot{}
\def\sectionmark##1{}\def\subsectionmark##1{}}
\renewcommand{\thefootnote}{\fnsymbol{footnote}}
\newcounter{sectionc}\newcounter{subsectionc}\newcounter{subsubsectionc}
\renewcommand{\section}[1] {\vspace*{0.6cm}\addtocounter{sectionc}{1}
\setcounter{subsectionc}{0}\setcounter{subsubsectionc}{0}\noindent
 {\normalsize\bf\thesectionc. #1}\par\vspace*{0.4cm}}
\renewcommand{\subsection}[1] {\vspace*{0.6cm}\addtocounter{subsectionc}{1}
        \setcounter{subsubsectionc}{0}\noindent
        {\normalsize\it\thesectionc.\thesubsectionc. #1}\par\vspace*{0.4cm}}
\renewcommand{\subsubsection}[1]
{\vspace*{0.6cm}\addtocounter{subsubsectionc}{1}\noindent{\normalsize\rm\thesectionc.\thesubsectionc.
\thesubsubsectionc. #1}\par\vspace*{0.4cm}}

\newcounter{appendixc}
\newcounter{subappendixc}[appendixc]
\newcounter{subsubappendixc}[subappendixc]

\renewcommand{\appendix}[1] {\vspace*{0.6cm}
        \refstepcounter{appendixc}
        \setcounter{figure}{0}
        \setcounter{table}{0}
        \setcounter{equation}{0}
        \renewcommand{\thefigure}{\Alph{appendixc}.\arabic{figure}}
        \renewcommand{\thetable}{\Alph{appendixc}.\arabic{table}}
        \renewcommand{\theappendixc}{\Alph{appendixc}}
        \renewcommand{\theequation}{\Alph{appendixc}.\arabic{equation}}
        \noindent{\bf Appendix \theappendixc #1}\par\vspace*{0.4cm}}

\def\abstracts#1{{\centering{\begin{minipage}{12.2truecm}\footnotesize\baselineskip=12pt\noindent
        \centerline{\footnotesize ABSTRACT}\vspace*{0.3cm}
        \parindent=0pt #1
        \end{minipage}}\par}}

\renewenvironment{thebibliography}[1]
	{\begin{list}{\arabic{enumi}.}
	{\usecounter{enumi}\setlength{\parsep}{0pt}
\setlength{\leftmargin 1.25cm}{\rightmargin 0pt}
\setlength{\itemsep}{0pt} \settowidth{\labelwidth}{88.}\sloppy}}{\end{list}}
\newcounter{itemlistc}
\newcounter{romanlistc}
\newcounter{alphlistc}
\newcounter{arabiclistc}

\newcommand{\fcaption}[1]{
        \refstepcounter{figure}
        \setbox\@tempboxa = \hbox{\footnotesize Fig.~\thefigure. #1}
        \ifdim \wd\@tempboxa > 6in
           {\begin{center}
        \parbox{6in}{\footnotesize\baselineskip=12pt Fig.~\thefigure. #1}
            \end{center}}
         \else
             {\begin{center}
             {\footnotesize Fig.~\thefigure. #1}
              \end{center}}
        \fi}
\newcommand{\tcaption}[1]{
        \refstepcounter{table}
        \setbox\@tempboxa = \hbox{\footnotesize Table~\thetable. #1}
        \ifdim \wd\@tempboxa > 6in
           {\begin{center}
        \parbox{6in}{\footnotesize\baselineskip=12pt Table~\thetable. #1}
            \end{center}}
        \else
             {\begin{center}
             {\footnotesize Table~\thetable. #1}
              \end{center}}
        \fi}
\def\@citex[#1]#2{\if@filesw\immediate\write\@auxout
        {\string\citation{#2}}\fi
\def\@citea{}\@cite{\@for\@citeb:=#2\do
        {\@citea\def\@citea{,}\@ifundefined
        {b@\@citeb}{{\bf ?}\@warning
        {Citation `\@citeb' on page \thepage \space undefined}}
        {\csname b@\@citeb\endcsname}}}{#1}}
\newif\if@cghi
\def\cite{\@cghitrue\@ifnextchar[{\@tempswatrue
 \@citex}{\@tempswafalse\@citex[]}}
\def\citelow{\@cghifalse\@ifnextchar [{\@tempswatrue
 \@citex}{\@tempswafalse\@citex[]}}
\def\@cite#1#2{{$\null^{#1}$\if@tempswa\typeout
        {IJCGA warning: optional citation argument
        ignored: `#2'} \fi}}

 1
 1
 1

\font\ninerm=cmr9

\begin{document}

\hspace*{\fill} IASSNS--HEP--95/55\\
\hspace*{\fill} hep--th/9506197 \\

\vspace*{1.5cm}

\centerline{\normalsize\bf REMARKS ON CHIRAL SYMMETRY BREAKING}
\baselineskip=16pt
\centerline{\normalsize\bf WITH MASSLESS FERMIONS}
\vspace*{0.9cm}

\centerline{\footnotesize RAINER DICK}
\baselineskip=15pt
\centerline{\footnotesize\it School of Natural Sciences, Institute
for Advanced Study}
\baselineskip=12pt
\centerline{\footnotesize\it Olden Lane, Princeton, NJ 08540, USA}
\centerline{\footnotesize and}
\centerline{\footnotesize\it Sektion Physik der Universit\"at M\"unchen}
\baselineskip=12pt
\centerline{\footnotesize\it Theresienstr.\ 37, 80333 M\"unchen,
Germany}
\vspace*{1.1cm}

\abstracts{In this talk I present recent results on Lorentz covariant
correlation functions $\langle q(p_1)\overline{q}(p_2)\rangle$
on the cone $p^2=0$.
In particular, chiral symmetry breaking terms are constructed
which resemble fermionic 2--point functions of 2--D CFT up to a scalar factor.}
\vfill
\noindent
{\footnotesize Talk presented at the International Workshop on Particle
Theory and Phenomenology, Ames (Iowa), May 1995.}
\newpage
\normalsize\baselineskip=15pt
\setcounter{footnote}{0}
\renewcommand{\thefootnote}{\alph{footnote}}
\noindent
{\bf 1.} Confinement and chiral symmetry breaking are characteristic
features of low energy QCD which remain puzzling from a theoretical
point of view, although there has been continuing progress
in the description of these phenomena during the
more than 20 years since the introduction of QCD.
Theoretical attempts
to understand confinement of relativistic quarks and gluons
from first principles center around the dual Meissner effect via
monopole condensation\cite{tho,man}.
Spontaneous breaking of chiral $SU(N_f)$ symmetry, on the other hand,
is expected to arise as
a consequence of confinement or as an instanton
effect\cite{CDG,evs}, but it is not clear which mechanism
drives chiral symmetry breaking.
Chiral symmetry breaking effects of confining forces have been
discussed in\cite{BC,corn}, and it has been pointed out that
in dual QCD
monopole condensation not only yields confinement but also
a chiral condensate through a gap equation\cite{BBZ}. Furthermore,
Seiberg and Witten recently constructed the Wilsonian
low energy effective action for $N=2$ super--Yang--Mills coupled
to hypermultiplets and determined the singularities of the
quantum moduli space\cite{SW}, see also\cite{AD}. They observed that
in this framework flavor charged monopoles imply both confinement
and chiral symmetry breaking.

In QCD we would like to understand
how ordinary monopoles break chiral symmetry in spite of their
chiral coupling, or whether
instantons or other non--perturbative effects break chiral
symmetry before confinement.
This problem is also relevant for the nature
of the phase transition, since the absence of an
order parameter for confinement
in the presence of light flavors
excludes a second order phase transition, if
chiral symmetry breaking is causally connected to
confinement\footnote{Without dynamical quarks the deconfinement transition
is expected to be first order from triality\cite{SY}.}.
Pisarski and Wilczek pointed out that an $\epsilon$--expansion
for the corresponding $\sigma$--model indicates a first order
transition
for more than two light flavors, but that a second order transition
in the universality class of the $O(4)$ vector model
is likely to appear in case of two light flavors\cite{PW,fw}.
A very recent review and discussion of numerical results can be found
in Ref.\cite{detar}.

In this talk I report on recent results for
the the correlation $\langle q(p_1)\overline{q}(p_2)\rangle$
on the orbit $p^2=0$, in an attempt to shed new light on the problem
from an unexpected angle\cite{rd1}.
Dynamical breaking of chiral flavor symmetry in gauge theories is a puzzle,
because it is very different from spontaneous magnetization:
In a ferromagnet the interaction tends to align the dipoles, while
thermal fluctuations restore disorder if the system has enough energy. Gluon
exchange, on the other hand, does not necessarily
align left-- and right--handed
fermions.
Stated differently, the massless Dyson--Schwinger equation for
the trace part of the fermion propagator always admits a
trivial solution, if a quark condensate is not inserted
{\it ab initio}\footnote{In the latter case the corresponding Dyson--Schwinger
equation yields a gap equation for the condensate. The puzzle then concerns
the identification of the "phonons" in QCD.}.
Therefore, chiral symmetry breaking has to be implemented
in unusual ways, if we want to recover it from
gauge dynamics: By requiring a
condensate as part of initial conditions,
in double scaling limits, through chiral symmetry breaking
boundary conditions, in chiral symmetry breaking regularizations, etc.
While this does not invalidate standard approaches to the problem, it serves
to remind the reader that some poorly understood mechanism leaves its
footprint on the long distance properties of the QCD vacuum, and
motivates the group theoretical
construction of Lorentz covariant correlation functions
given below.

The main ingredient of the work reported below is a mapping
between massless spinors in 3+1 dimensions and primary fields,
which relates the order parameter
to automorphic forms
under the Lorentz group. From a mathematical point of view,
the novel feature of the automorphic forms under
investigation is that they provide correlations between primary fields
on spheres of different radii, thus providing true representations of
the Lorentz group and extending the determination of correlation functions
in 2D conformal field theory. The nontrivial behavior of radii under the
boost sector of the Lorentz group allows for chiral symmetry preserving
terms in the correlation
functions which could not appear in a two--dimensional framework, while the
chiral symmetry breaking terms in turn appear
closely related to 2D fermionic correlation
functions.

I will briefly review the evidence for
dynamical chiral symmetry breaking in QCD in Sec.\ 2. The mapping between
massless spinors in Minkowski space and primary fields on spheres in
momentum space and an application to construct chiral
symmetry breaking correlations will be outlined in Sec.\ 3.
\\[2ex]
{\bf 2.} There exists wide agreement that chiral symmetry
breaking in QCD arises both dynamically, as a genuine QCD phenomenon, and
through electroweak symmetry breaking, which in a standard scenario accounts
for the quark current masses\footnote{The electroweak sector
also contributes to breaking of chiral $SU(N_f)$ through the axial anomaly
since the charge operator $Q^2$ is not flavor symmetric.}.
Dynamical chiral symmetry breaking is then expected
to account partially for the difference
between current and constituent masses\cite{pol}. From this point of view,
the large discrepancy between current and constituent
masses, and the fact that there is not even an approximate parity degeneracy
in the hadron spectrum provides strong evidence for dynamical breaking of
chiral
symmetry.

Another argument in favor of dynamical breaking
of chiral symmetry comes from the Gell-Mann--Oakes--Renner relation:
\begin{equation}
2m_q\langle \overline{q}q\rangle = -f_{\pi}^2 m_{\pi}^2
\end{equation}
where $m_q$ stands for a mean value of current quark masses.
This relation is expected to hold in the sense of a leading approximation
in $m_q$, and works phenomenologically
the better the smaller the value of $m_q$
is\cite{DGH,leut}.
While this does not strictly imply $\lim_{m_q\to 0}\langle \overline{q}q\rangle
\neq 0$, it implies at least that the condensate vanishes weaker than first
order
in $m_q$.

The necessity of including non--vanishing condensates in QCD sum rules
provides further strong indication for spontaneous breaking of chiral symmetry.
This becomes particularly evident in heavy--light systems, where the
condensate of the light quark is expected
to contribute to the meson propagator even
in the limit of vanishing current mass\cite{shif,RRY}.

 Further hints for
chiral symmetry breaking are provided by
`t Hooft's result that decoupling of heavy fermions does not comply
with local chiral flavor symmetry\cite{tH}, and indirectly
through the no--go theorem of Vafa and Witten
for spontaneous breaking
of vector--like global symmetries in QCD\cite{VW}.

Chiral symmetry breaking can also be addressed
in lattice QCD with staggered
fermions. In this framework
non--vanishing condensates have been
reported e.g.\ in\cite{bow,born,sal,gupta}.

This mini--review comprises a very short summary of the
most compelling arguments in favor of dynamical chiral symmetry breaking.
Clearly, there is insurmountable evidence that
chiral symmetry breaking in QCD is not solely of electroweak
origin.\\[2ex]
{\bf 3.} Chiral spinors in 3+1 dimensions can be described as
primary fields of conformal weight $\frac{1}{2}$
on spheres in momentum space.
To exploit this observation, we work in
the Weyl representation of Dirac matrices,
and parametrize the unit sphere in momentum space in terms of
stereographic coordinates:
\begin{equation}\label{zdef1}
z=\frac{p_1+ip_2}{|{\bf p}|-p_3}\qquad\qquad\tilde{z}=
-\frac{p_1-ip_2}{|{\bf p}|+p_3}
\end{equation}
Proper orthochronous Lorentz transformations act on these coordinates
according to
\begin{equation}\label{zlor1}
z^{\prime}=z({\bf p}^{\prime})=\overline{U}\circ z({\bf p})=
\frac{\bar{a}z+\bar{b}}{\bar{c}z+\bar{d}}
\end{equation}
if $E=|{\bf p}|$, and
\begin{equation}\label{zlor2}
z^{\prime}=U^{-1T}\circ z({\bf p})=
\frac{dz-c}{a-bz}
\end{equation}
if $E=-|{\bf p}|$.\\
$U$ denotes the positive chirality spin $\frac{1}{2}$
representation of the Lorentz
group:
\[
U(\omega)=\exp(\frac{1}{2}\omega^{\mu\nu}\sigma_{\mu\nu}^{})
=\left(\begin{array}{cc} a & b\\ c & d\end{array}\right)\in
SL(2,\mbox{\bf C})
\]
We identify local functions written in co-ordinates
$(z,\bar{z},|\bf p|)$ and
$(\tilde{z},\bar{\tilde{z}},|{\bf p}|)$ via
\begin{equation}\label{weyl}
\psi(\tilde{z},\bar{\tilde{z}},|{\bf p}|)= -z\psi(z,\bar{z},|{\bf p}|)
\end{equation}
\begin{equation}\label{antiweyl}
\phi(\tilde{z},\bar{\tilde{z}},|{\bf p}|)= \bar{z}\phi(z,\bar{z},|{\bf p}|)
\end{equation}
and these overlap conditions can be rephrased as Weyl equations:
\[
(|{\bf p}|+{\bf p}\cdot{\bf\sigma})
\left(\begin{array}{c}\psi(z,\bar{z},|{\bf p}|)\\
\psi(\tilde{z},\bar{\tilde{z}},|{\bf p}|)\end{array}\right)=0
\]
\[
(|{\bf p}|-{\bf p}\cdot{\bf\sigma})\left(\begin{array}{c}
\phi(\tilde{z},\bar{\tilde{z}},|{\bf p}|)\\ \phi(z,\bar{z},|{\bf p}|)
\end{array}\right)=0
\]
Under (\ref{zlor1}) $\phi$ and $\psi$ transform according to
\begin{equation}\label{traphi}
\phi^{\prime}(z^{\prime},\bar{z}^{\prime},|{\bf p}^{\prime}|)=
(c\bar{z}+d)\phi(z,\bar{z},|{\bf p}|)
\end{equation}
\begin{equation}\label{trapsi}
\psi^{\prime}(z^{\prime},\bar{z}^{\prime},|{\bf p}^{\prime}|)=
(\bar{c}z+\bar{d})\psi(z,\bar{z},|{\bf p}|)
\end{equation}
if $E=|{\bf p}|$.
Due to (\ref{weyl},\ref{antiweyl}) this is equivalent to
\[
\left(\begin{array}{c}
\phi^{\prime}(\tilde{z}^{\prime},\bar{\tilde{z}}^{\prime},|{\bf p}^{\prime}|)\\
\phi^{\prime}(z^{\prime},\bar{z}^{\prime},|{\bf p}^{\prime}|)\end{array}\right)
=U\cdot
\left(\begin{array}{c}\phi(\tilde{z},\bar{\tilde{z}},|{\bf p}|)\\
\phi(z,\bar{z},|{\bf p}|)\end{array}\right)
\]
\[
\left(\begin{array}{c}
\psi^{\prime}(z^{\prime},\bar{z}^{\prime},|{\bf p}^{\prime}|)\\
\psi^{\prime}(\tilde{z}^{\prime},\bar{\tilde{z}}^{\prime},
|{\bf p}^{\prime}|)\end{array}\right)
=U^{-1\dagger}\cdot
\left(\begin{array}{c}\psi(z,\bar{z},|{\bf p}|)\\
\psi(\tilde{z},\bar{\tilde{z}},|{\bf p}|)\end{array}\right)
\]
The case $E=-|{\bf p}|$ corresponds to
$U\leftrightarrow U^{-1\dagger}$ in the equations above, but we will stick
to positive energy in the sequel. Correlations for $E=-|{\bf p}|$
can easily be
recovered
from the covariance cosiderations for positive energy through
a reflection ${\bf p}\to -{\bf p}$.

To make a short story even shorter, we
made spin bundles look like line bundles over the orbit $p^2=0$.
To take advantage of this construction, we write a spinor on the
half--cone $E=|{\bf p}|$:
\begin{equation}\label{expand}
\Psi(p)=\left(\begin{array}{c}1\\0
\end{array}\right)
\otimes \left(\begin{array}{c}\bar{z}\\1\end{array}\right)
\phi(z,\bar{z},|{\bf p}|)+
\left(\begin{array}{c}0\\1
\end{array}\right)
\otimes \left(\begin{array}{c}1\\-z\end{array}\right)\psi(z,\bar{z},|{\bf p}|)
\end{equation}
with a corresponding representation of the correlation
function of massless fermions in the Dirac picture
\begin{equation}\label{prop}
\langle\Psi(p)\overline{\Psi}(p^{\prime})\rangle=
\end{equation}
\[
\left(\begin{array}{cc}0&1\\0&0\end{array}\right)\otimes
\left(\begin{array}{cc}\bar{z}z^\prime & \bar{z}\\
z^\prime&1\end{array}\right)\langle\phi({\bf p})\phi^+({\bf
p}^{\prime})\rangle+
\left(\begin{array}{cc}0&0\\1&0\end{array}\right)\otimes
\left(\begin{array}{cc}1 & -\bar{z}^\prime\\
-z& z\bar{z}^\prime
\end{array}\right)\langle\psi({\bf p})\psi^+({\bf p}^{\prime})\rangle
\]
\[
+
\left(\begin{array}{cc}1&0\\0&0\end{array}\right)\otimes
\left(\begin{array}{cc}\bar{z} & -\bar{z}\bar{z}^\prime\\
1&-\bar{z}^\prime
\end{array}\right)\langle\phi({\bf p})\psi^+({\bf p}^{\prime})\rangle +
\left(\begin{array}{cc}0&0\\0&1\end{array}\right)\otimes
\left(\begin{array}{cc}z^\prime & 1\\
-zz^\prime&-z\end{array}\right)\langle\psi({\bf p})\phi^+({\bf p}^{\prime})
\rangle
\]
The 2--point functions on the right hand side
transform under a factorized
representation of the Lorentz group.
This makes this representation very convenient for the investigation
of correlations $\langle\Psi(p)\overline{\Psi}(p^\prime)\rangle$
which comply with Lorentz covariance.
Stated differently, we ask which correlations
of spinors of the form  $\Psi(p)$ could be constructed
from a non--trivial vacuum, or more general, between any Lorentz invariant
states.

The investigation in Ref.\cite{rd1} revealed
\begin{equation}\label{f1}
\langle\psi({\bf p}_1)\psi^+({\bf p}_2)\rangle=
\langle\phi({\bf p}_2)\phi^+({\bf p}_1)\rangle=
f_1\!\left(\frac{|{\bf p}_1|}{|{\bf p}_2|}\right)
\frac{1+z_1\bar{z}_2}{\sqrt{|{\bf p}_1||{\bf p}_2|}}\,
\delta_{z\bar{z}}(z_1-z_2)
\end{equation}
\begin{equation}\label{f2}
\langle\psi({\bf p}_1)\phi^+({\bf p}_2)\rangle=
\overline{\langle\phi({\bf p}_2)\psi^+({\bf p}_1)\rangle}=
\frac{1}{z_1-z_2}\,
f_2\!\left(|{\bf p}_1||{\bf p}_2|
\frac{(z_1-z_2)(\bar{z}_1-\bar{z}_2)}
{(1+z_1\bar{z}_1)(1+z_2\bar{z}_2)}\right)
\end{equation}
where Lorentz covariance does not fix $f_1$ and
$f_2$\footnote{The correlation in the vacuum of the free theory
$
\langle\psi({\bf p})\overline{\psi}({\bf p}^{\prime})\rangle =
-2p\cdot\gamma|{\bf p}|\delta({\bf p}-{\bf p}^{\prime})
$
is recovered from Eqs.\ (\ref{prop},\ref{f1},\ref{f2}) for
$f_1(x)=\delta(x-1)$, $f_2=0$.}.
However, on dimensional grounds we infer that
$f_2(x)=\frac{C}{\sqrt{x}}$, whence the correlation function is strongly
peaked for parallel momenta\footnote{Inclusion of a scale $\Lambda$
allows for obvious generalizations.}.
Under the proviso that
a Dirac picture makes sense in a
nonperturbative problem,
the orbit $p^2=0$ contributes to a condensate
\begin{equation}\label{cond}
tr\langle\Psi(p)\overline{\Psi}(p^{\prime})\rangle=
-\frac{2C}{\sqrt{|{\bf p}||{\bf p}^{\prime}|}\sin(\frac{\theta}{2})}
\end{equation}
where $\theta$ denotes the angle between ${\bf p}$ and
${\bf p}^{\prime}$.

If the correlation function in configuration space
gives the positive energy contribution to a propagator
of initial conditions (modulo $i\gamma_0$), as follows for the
perturbative vacuum from
canonical quantization, then the result above would be consistent
insofar as
the $f_2$ terms do not
anticommute with $\gamma_5^{}$, while the $f_1$ terms
anticommute with $\gamma_5^{}$ and imply a restriction for external
momenta to be parallel.
However, since we pretend to deal with a confining theory (albeit disguised
in a non--perturbative vacuum), there is no reason to believe that
the propagator can be reconstructed from data on a single orbit
of the Lorentz group.
The results above presumably make sense only within a confining
theory, if observables are expressed in meson or
baryon correlations.
Of course, $n$--point functions of primary fields in 2D CFT are not
unique if $n>3$, and I expect similar ambiguities to show up in the present
setting.\\[2ex]
{\bf Acknowledgements}:
I would like to thank Frank Wilczek for helpful comments about order
parameters in QCD. This work was supported by the DFG.\\[2ex]

\end{document}